\documentclass[12pt,letterpaper,notitlepage]{article}

\usepackage{epsfig,amsmath,amssymb,cite,color,multirow,ulem}

\newcommand{\tev}{\,\, \mathrm{TeV}}
\newcommand{\gev}{\,\, \mathrm{GeV}}

\begin{document}

\begin{titlepage}

\begin{flushright}
PITT-PACC-1404
\end{flushright}

\vspace{15pt}
\begin{center}
\LARGE Leptophilic Dark Matter in Lepton Interactions\newline at LEP and ILC
\end{center}

\vspace{0pt}
\begin{center}
{\large A.~Freitas and S.~Westhoff}\\
\vspace{30pt} {
PITTsburgh Particle-physics Astro-physics \& Cosmology
    Center (PITT-P\"ACC),\newline  Department of Physics \& Astronomy,
    University of Pittsburgh, Pittsburgh, PA 15260, USA
   }
\end{center}

\vspace{10pt}
\begin{abstract}
\vspace{2pt} 
\noindent
Dark matter particles that couple primarily to leptons are only weakly
constrained by data from direct detection experiments and the LHC. However,
models of this kind necessarily generate effective four-lepton contact
interactions at the tree- or one-loop-level, which can be probed in $e^+e^-$
collisions. In this work, precise data from LEP is used to derive limits on
leptophilic dark matter in a model-independent framework. The bounds turn out to
be competitive with, and in some cases exceed, limits from mono-photon events.
We also extrapolate our results to ILC energies and luminosities. We show that
ILC data for contact interactions may be able to set the strongest limits on
TeV-scale leptophilic dark matter for couplings $g\gtrsim 1$ of the mediators.
\end{abstract}

\end{titlepage}

\clearpage


\section{Introduction}

Weakly interacting dark matter (DM) particles are actively being searched for
 through a variety of methods. Strong bounds on their parameter space have been established by direct detection \cite{dd}, indirect detection \cite{id} and collider \cite{lhcd} experiments. However, these limits are weakened substantially in leptophilic DM models, see for example Refs.~\cite{leptophil1, leptophil2}. Such models feature weak-strength
 interactions between the DM particle, $\chi$, and Standard Model (SM) leptons  at the tree level, but DM--quark couplings are induced only at the loop level \cite{leptophil1,Dreiner:2012xm}. Some of the strongest collider bounds on leptophilic models thus stem from searches for mono-photon events at $e^+e^-$ machines.
This signature is characteristic for the process $e^+e^- \to \chi\chi\gamma$, where an initial-state photon recoils against the DM particle pair, which is not observed in the detector.

Constraints on leptophilic DM from mono-photon searches at LEP have been analyzed in Ref.~\cite{Fox:2011fx}. These limits may be greatly improved by a future International Linear Collider (ILC) with a center-of-mass (CM) energy of $500\gev$ -- $1\tev$ \cite{Dreiner:2012xm,ilc2}. The precise bounds depend on the mass, spin and couplings of the DM particle and the mediator that communicates the interaction between DM and SM particles.
In this work, however, it is shown that comparable or even stronger bounds can be obtained from the analysis of four-lepton contact interactions contributing to the process $e^+e^-\to \ell^+\ell^-$, where $\ell=e,\mu, \tau$. LEP has put tight constraints on any new-physics contributions to these four-lepton processes \cite{lep2}, and these limits are expected to be substantially strengthened at the ILC. We here demonstrate that in leptophilic DM scenarios, such four-lepton contact interactions are necessarily generated either through tree-level exchange of the mediator or through loop contributions involving the  DM and mediator particles.

We use a model-independent framework where the SM is extended by a single DM field and a single mediator field, with arbitrary renormalizable
couplings and arbitrary spin (up to spin 1) of the new particles
\cite{Dreiner:2012xm}. This framework will be introduced in section~\ref{models}. The four-fermion bounds are derived in an effective field
theory (EFT) framework, where it is assumed that the mediator particle is heavy compared to the $e^+e^-$ CM energy. Limits derived from LEP data will be presented in section~\ref{tree} and section~\ref{loop} for scenarios where the four-lepton interaction is generated at tree-level or loop-level, respectively. In section~\ref{ilc}, we use these results to estimate the projected reach of the ILC for leptophilic DM through contact interactions. Finally, comparisons with non-collider experiments and the applicability of the EFT are discussed in section~\ref{disc}, before concluding in section~\ref{concl}.


\section{Models}
\label{models}

To define our framework of leptophilic DM models, we adopt the classification of simplified models from Ref.~\cite{Dreiner:2012xm}. Only renormalizable models are considered, which
implies that the interaction between the DM and SM leptons is facilitated by a
(heavy) mediator particle. The models are characterized by the spins (0,
$\frac{1}{2}$ or 1) of the DM particle (denoted $\chi$) and of the mediator
particle (denoted $\eta$). The dark/hidden sector may contain additional heavy particles, but it is assumed that they are not relevant for the DM phenomenology. To ensure its stability,
$\chi$ is assigned to be odd under a $\mathbb{Z}_2$ symmetry, while $\eta$ is $\mathbb{Z}_2$-even or -odd depending on the form of the interaction.

In Tab.~\ref{tab:mod}, the full list of models and the form of their interactions is given. Here the spins of $\chi$ and $\eta$ are denoted by the letters ``S'', ``F'' and ``V'' for scalar, fermion and vector, respectively. For a bosonic mediator $\eta$, one can construct models where $\eta$ appears either in the $s$- or $t$-channel of the annihilation diagram $\chi\chi \to \ell^+\ell^-$. To distinguish these possibilities, the $t$-channel mediation is denoted by ``tS'' or ``tV''.

\begin{table}[tb]
\centering
\renewcommand{\arraystretch}{1.5}
\begin{tabular}{|cc|l|l|}
\hline
DM ($\chi$) & Med.\  ($\eta$) & Coupling to SM ($-{\cal L}_{\rm int}$) & Benchmark scenarios \\
\hline\hline
S & S & $g'\chi^\dagger\chi\eta + \bar{\psi}(g_s+g_p\gamma_5)\psi\eta$ &
 \multirow{3}{*}{\parbox{11.5em}{$g_p=g'_p=0$ (scalar)\\[1ex] $g_s=g'_s=0$ (pseudoscalar)}} \\
\cline{1-3}
F & S & $\bar{\chi}(g'_s+g'_p\gamma_5)\chi\eta +
 \bar{\psi}(g_s+g_p\gamma_5)\psi\eta$ & \\
\cline{1-3}
V & S & $g'\chi_\mu\chi^\mu\eta + \bar{\psi}(g_s+g_p\gamma_5)\psi\eta$ & \\
\hline\hline
S & V & $g'\chi^\dagger \stackrel{\leftrightarrow}{\partial}_\mu \chi \eta^\mu +
 \bar{\psi}\gamma_\mu(g_v+g_a\gamma_5)\psi\eta^\mu$ &  
 \multirow{3}{*}{\parbox{11.5em}{$g_a=g'_a=0$ (vector)\\[1ex] $g_v=g'_v=0$ (axialvector)}} \\
\cline{1-3}
F & V & $\bar{\chi}\gamma_\mu(g'_v+g'_a\gamma_5)\chi\eta^\mu +
 \bar{\psi}\gamma_\mu(g_v+g_a\gamma_5)\psi\eta^\mu$ & \\
\cline{1-3}
V & V & \parbox[c][3em]{15em}{$ig'(\eta_\mu\chi^\dagger_\nu\chi^{\mu\nu} +
 \chi_\mu\eta_\nu\chi^{\dagger\mu\nu} + \chi^\dagger_\mu\chi^\nu\eta^{\mu\nu})$
 \\[.5ex] ${}\; + \bar{\psi}\gamma_\mu(g_v+g_a\gamma_5)\psi\eta^\mu$} & \\
\hline\hline
S & F & $\bar{\eta}(g_lP_L + g_rP_R)\psi\chi + \text{h.c.}$ &   
 \multirow{2}{*}{\parbox{11.5em}{$g_l=0$ (right-handed)\\[1ex] $g_r=0$ (left-handed)}} \\
\cline{1-3}
V & F & $\bar{\eta}\gamma_\mu(g_lP_L + g_rP_R)\psi\chi^\mu + \text{h.c.}$ & \\
\hline\hline
F & tS & $\bar{\chi}(g_lP_L + g_rP_R)\psi\eta + \text{h.c.}$ &   
 \multirow{2}{*}{\parbox{11.5em}{$g_l=0$ (right-handed)\\[1ex] $g_r=0$ (left-handed)}} \\
\cline{1-3}
F & tV & $\bar{\chi}\gamma_\mu(g_lP_L + g_rP_R)\psi\eta^\mu + \text{h.c.}$ & \\
\hline
\end{tabular}
\caption{List of simplified models involving a DM field $\chi$, mediator field $\eta$ and SM lepton field $\psi$ \cite{Dreiner:2012xm}. Here ``S'', ``F'' and ``V'' denote fields of spin 0, $\frac{1}{2}$ and 1, respectively, while ``tS'' and ``tV'' indicate that the mediator is exchanged in the $t$-channel of the DM annihilation process. Moreover, $P_{L,R} = (1\pm\gamma_5)/2$ project onto chiral fermion states.}
\label{tab:mod}
\end{table}

To limit the size of the parameter space, we focus on specific benchmark scenarios for the couplings introduced in the middle column of Tab.~\ref{tab:mod}. For the interaction of a spin-0 (spin-1) $s$-channel mediator with fermions, we consider either pure scalar (vector) or pseudo-scalar (axial-vector) couplings. For the $t$-channel mediators, it is more natural to use left- or right-handed couplings instead, to ensure SU(2)-gauge invariance of the $\ell$-$\chi$-$\eta$ interaction. In addition, we assume that the mediator couples to the leptons of all three generations with the same strength, but its coupling to SM quarks is zero.  We further assume that the DM field is a lepton flavor singlet.

The models can be further distinguished depending on whether the DM field is real (self-conjugate) or complex ($i.\,e.$ with distinct particle and antiparticle components). As in Ref.~\cite{Dreiner:2012xm}, we denote the former case with a suffix ``r''. For example, ``FS'' denotes Dirac DM, while ``FSr'' indicates Majorana DM.

In the next two sections, the constraints from four-lepton interactions at LEP on these models will be analyzed. The program {\sc FeynArts} \cite{feynarts} has been used to generate the necessary one-loop amplitudes. We have performed the loop integration in two ways using {\sc FeynCalc} \cite{Mertig:1990an} and a private computer code to obtain a cross-check of our calculation.


\section{S-channel mediation: tree-level lepton interactions}
\label{tree}

In the first six models in Tab.~\ref{tab:mod}, the DM annihilation is mediated by the $s$-channel exchange of a scalar or vector boson. The same bosonic mediator will necessarily also lead to a tree-level contribution to the four-lepton processes $e^+e^- \to \ell^+\ell^-$. These are strongly constrained from measurements of four-lepton contact interactions \cite{lep2} and di-lepton resonance searches in $e^+e^- \to \ell^+\ell^-\gamma$ \cite{Abbiendi:1999wm}. Following the analysis in section~3 of Ref.~\cite{Freitas:2014pua}, we obtain the following bounds for a spin-1 mediator at 90\% C.L.
\begin{align}\label{eq:s-v}
&\text{vector:} & g_v/M_\eta &< 2.0\times 10^{-4} \gev^{-1} & &(M_\eta>200\gev), \\
& & g_v/M_\eta &< 6.9\times 10^{-4} \gev^{-1} & &(100\gev<M_\eta<200\gev), \\[1ex]
&\text{axial-vector:} & g_a/M_\eta &< 2.4\times 10^{-4} \gev^{-1} & &(M_\eta>200\gev), \\
& & g_a/M_\eta &< 6.9\times 10^{-4} \gev^{-1} & &(100\gev<M_\eta<200\gev).
\end{align}\label{eq:s-s}
Similar bounds have been obtained in Ref.~\cite{Bell:2014tta}. For a spin-0 mediator we find
\begin{align}
&\text{(pseudo)scalar:} & g_{s,p}/M_\eta &< 2.7\times 10^{-4} \gev^{-1} & &(M_\eta>200\gev), \\ & & g_{s,p}/M_\eta &< 7.3\times 10^{-4} \gev^{-1} & &(100\gev<M_\eta<200\gev).
\end{align}
These bounds exceed the limits from mono-photon searches at LEP in Ref.~\cite{Fox:2011fx} by about one order of magnitude. Furthermore, they are independent of the DM mass and therefore hold beyond the kinematic limit of direct pair production at LEP2, $M_{\chi}\lesssim 100\gev$. Currently four-lepton interactions at LEP thus provide the strongest constraints on leptophilic DM with $s$-channel mediation.


\section{T-channel mediation: loop-level lepton interactions}
\label{loop}

\begin{figure}[tb]
\centering
\includegraphics[width=13cm]{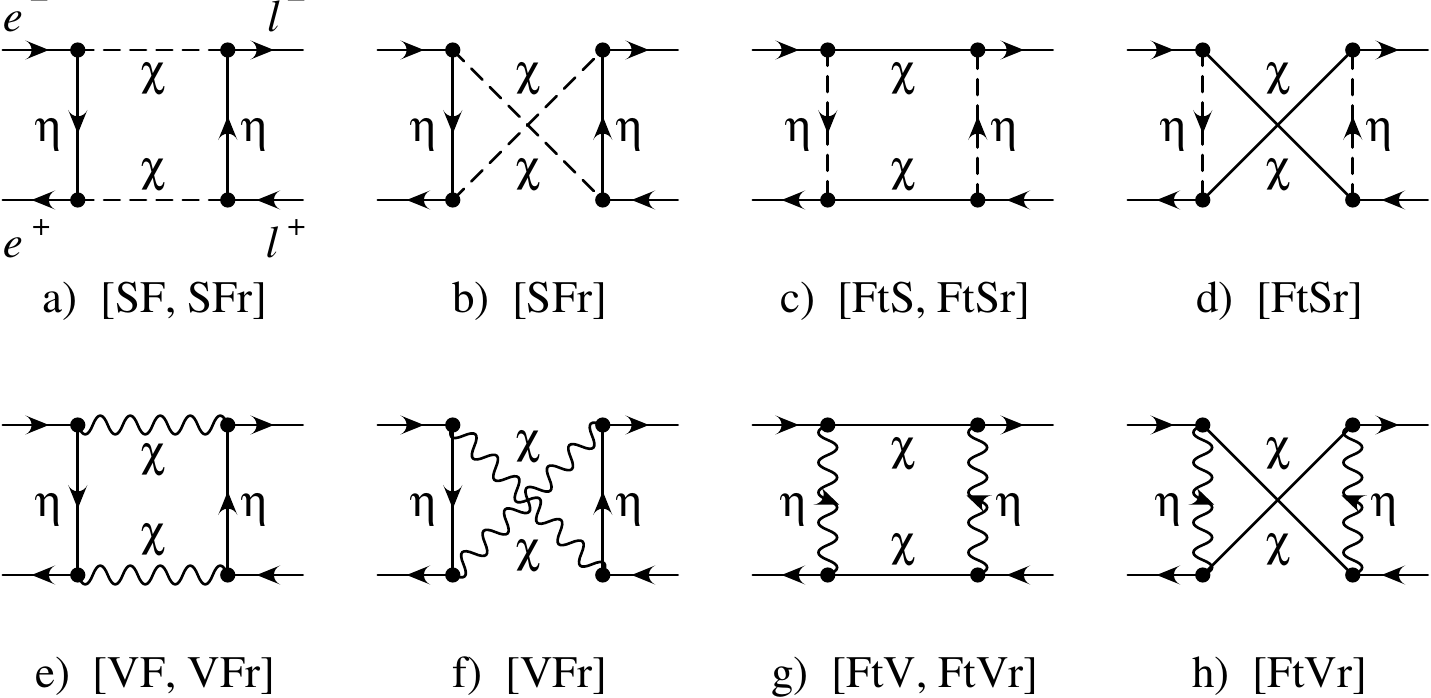}
\caption{Feynman diagrams for the one-loop box contributions to four-lepton contact interactions. The relevant model scenarios are given in square brackets.}
\label{fig:diags}
\end{figure}

The models with $t$-channel mediators considered in this work are defined in the last four rows in Tab.~\ref{tab:mod}. In these scenarios, $e^+ e^-\rightarrow \ell^+\ell^-$ transitions are generated at the one-loop level through the box diagrams in Fig.~\ref{fig:diags}. Here and in the following discussion we restrict ourselves to muons in the final state, i.e. to $\ell=\mu$. Owing to the loop suppression, constraints on $t$-channel models are expected to be weaker than those on $s$-channel models discussed in the previous section. But as we will see, they can be competitive with limits from direct DM production through $e^+e^- \to \chi\chi\gamma$. In fact, for large couplings $g_r > 1$, four-lepton interactions yield the strongest bounds on leptophilic DM scenarios with $t$-channel mediation.

Since the mediator carries electric charge, its mass should exceed $M_{\eta} \gtrsim 100\gev$ to evade constraints from direct pair production at LEP. It should furthermore be heavier than the DM particle, $M_{\eta} > M_{\chi}$, to ensure DM stability. If the mass of the particles in the loop is larger than the beam energy at LEP2, $M_{\eta,\chi} > 200\,\gev$, DM effects in $e^+e^-\rightarrow \ell^+\ell^-$ can be described by effective four-lepton interactions 
\begin{align}
\mathcal{H}_{\rm eff} & = \sum_{A} \mathcal{C}_{A}\,\mathcal{O}_{A}\,.
\end{align}
Since we are considering chiral interactions, the only relevant local operators are
\begin{align}\label{eq:ops}
\mathcal{O}_{LL} & = (\overline{e}\gamma_{\mu}P_Le)(\overline{\ell}\gamma^{\mu}P_L\ell)\,, & \mathcal{O}_{RR} & = (\overline{e}\gamma_{\mu}P_Re)(\overline{\ell}\gamma^{\mu}P_R\ell)\,,
\end{align}
and corresponding Wilson coefficients $\mathcal{C}_{LL}$ and $\mathcal{C}_{RR}$. In scenarios with Majorana fermion DM, the diagrams in Fig.~\ref{fig:diags} (d) and (h) also introduce the scalar operators $\mathcal{O}_{RL} = (\overline{\ell} P_R e)(\overline{\ell}P_L e)$ and $\mathcal{O}_{LR} = (\overline{\ell} P_L e)(\overline{\ell}P_R e)$. Through a Fierz transformation, $\mathcal{O}_{RL}$ and $\mathcal{O}_{LR}$ can be mapped onto the vector operators in (\ref{eq:ops}), yielding $\mathcal{O}_{RL} = \mathcal{O}_{RR}/2$ and $\mathcal{O}_{LR} = \mathcal{O}_{LL}/2$, respectively.

If the DM particle is light, $M_\chi < 200\gev < M_{\eta}$, it should in principle remain in the spectrum of the EFT. As a consequence, there will be additional operators $\ell\ell\chi\chi$ (here $\ell=e,\mu,\tau$), mediated by the heavy $\eta$, which mix into the four-lepton operators. Here it is assumed that this mixing contribution is small, and thus it is sufficient to focus on the effective four-lepton operators in \eqref{eq:ops}. This assumption is in general not expected to be a good approximation for the imaginary part of the amplitude ${\cal A}_{\rm NP}$, due to the existence of a physical cut for $M_\chi < 100\gev$. However, since the leading contribution of the contact interactions is given by the interference term with the SM amplitude ${\cal A}_{\rm SM}$, $2\,\text{Re}\{{\cal A}^*_{\rm SM}\,{\cal A}_{\rm NP}\}$, and the imaginary part of ${\cal A}_{\rm SM}$ is small, the effect on the observable cross-section can be neglected.

Special care is required for the cases where either the DM particle or the mediator is a massive vector boson. The scenario VFr contains a neutral real vector particle, which can be regarded as a (fundamental) gauge boson that receives mass through spontaneous symmetry breaking. This can be verified explicitly by computing the box diagrams in Fig.~\ref{fig:diags}~(e,f) in a general $R_\xi$ gauge, including contributions from Goldstone boson exchange, and checking that the result is independent of the gauge parameter $\xi$.

On the other hand, the scenarios VF, FtV and FtVr contain a \textit{complex} vector boson of mass $M_V$, with distinct particle and antiparticle components. Such a complex vector particle can occur as a mesonic bound state of strong dynamics with a fundamental scale $\Lambda_c$. In this case, however, the description of the vector meson with a basic vector propagator
\begin{align}
\frac{-ig_{\mu\nu}+ik_\mu k_\nu/M_V^2}{k^2-M_V^2} \label{eq:prop}
\end{align}
will break down for energies $k^2 \gtrsim \Lambda_c^2$. In fact, the loop integrals for the four-lepton box diagrams in Fig.~\ref{fig:diags}~(e,g,h) are UV-divergent for the scenarios VF, FtV and FtVr when using the propagator \eqref{eq:prop}. The UV singularity can be cured by multiplying the propagator with a form factor $F(k^2,\Lambda_c)$. The precise form of $F(k^2,\Lambda_c)$ is model-dependent and difficult to derive from first principles. This lack of knowledge about $F(k^2,\Lambda_c)$ introduces an unavoidable source of theoretical uncertainty in the EFT. A simple choice with the proper high-energy behavior is given by \cite{Rizzo:1984bc}
\begin{align}
\biggl ( \frac{-ig_{\mu\nu}+ik_\mu k_\nu/M_V^2}{k^2-M_V^2} \biggr )
\frac{1}{1-k^2/\Lambda_c^2}. \label{eq:form}
\end{align}
As a result of using this modified propagator, the results for the one-loop induced four-lepton interactions for VF, FtV and FtVr will depend on the additional scale $\Lambda_c$.
Note that this sensitivity of the low-energy EFT to the high-scale dynamics is also visible in the divergent behavior of the mono-photon cross-section for vector DM, $e^+e^- \to \chi_V\chi_V\gamma$, in the limit $M_{\chi_V} \to 0$ and $\Lambda_c \to \infty$ \cite{Dreiner:2012xm}.

The one-loop four-lepton interactions for the benchmark scenarios in Tab.~\ref{tab:mod} are given in Tab.~\ref{tab:4lci}.
\begin{table}[tb]
\centering
\renewcommand{\arraystretch}{1.5}
\begin{tabular}{|l|l|}
\hline
Model & Effective interaction $\mathcal{C}_{RR}\,\mathcal{O}_{RR}$\\
\hline
\hline
SF & $\frac{g_r^4}{64\pi^2M_\eta^2}F_{\rm S}(x)\,\mathcal{O}_{RR}$ \\
\hline
SFr & 0 \\
\hline
FtS & $-\frac{g_r^4}{64\pi^2M_\eta^2}F_{\rm S}(x)\,\mathcal{O}_{RR}$ \\
\hline
FtSr & $-\frac{g_r^4}{64\pi^2M_\eta^2}\big[F_{\rm S}(x) + F_{\rm Sr}(x)\big]\mathcal{O}_{RR}$ \\
\hline
VF & $\frac{g_r^4}{64\pi^2M_\eta^2}\,\frac{\Lambda_c^2}{M_{\chi}^2}\,F_{\rm V}(x,y)\,\mathcal{O}_{RR}$ \\
\hline
VFr & $-\frac{3g_r^4}{16\pi^2M_\eta^2}F_{\rm S}(x)\,\mathcal{O}_{RR}$ \\
\hline
FtV & $-\frac{g_r^4}{64\pi^2M_\eta^2}\,\frac{\Lambda_c^2}{M_{\chi}^2}\,F_{\rm V}(1/x,y/x)\,\mathcal{O}_{RR}$\\
\hline
FtVr & $-\frac{g_r^4}{64\pi^2M_\eta^2}\,\frac{\Lambda_c^2}{M_{\chi}^2}\big[F_{\rm V}(1/x,y/x) + F_{\rm Vr}(1/x,y/x)\big]\,\mathcal{O}_{RR}$\\
\hline
\end{tabular}
\caption{Effective four-lepton interactions for DM scenarios with heavy $t$-channel mediator fields $\eta$ and right-handed couplings $g_r$. The results with left-handed couplings are obtained by the replacement $g_r \to g_l$ and $\mathcal{O}_{RR}\to\mathcal{O}_{LL}$. The loop functions $F_{\rm S,Sr}(x)$ and $F_{\rm V,Vr}(x,y)$ with $x = M^2_{\chi}/M_{\eta}^2$ and $y = M_{\chi}^2/\Lambda_c^2$ are defined in (\ref{eq:ciloop}). Results with complex vector bosons depend on the compositeness scale $\Lambda_c$.}
\label{tab:4lci}
\end{table}
 We have defined the loop functions
\begin{align}\label{eq:ciloop}
F_{\rm S}(x) &= \tfrac{1}{(1-x)^3} \bigl [1 - x^2 + 2x\ln x \bigr],\\\nonumber
F_{\rm Sr}(x) &= \tfrac{2x}{(1-x)^3} \bigl [2 - 2x + (x+1)\ln x \bigr],\\[1ex]\nonumber
F_{\rm V}(x,y) &= \tfrac{1}{(1-x)^3(1-y)^3(x-y)^3} \bigl \{\bigl [
 x^2(1-x)(1-y)(1-2x+x^2-14y+19xy-11x^2y+xy^2\\\nonumber
&\hspace*{2cm}+4x^2y^2) - 2xy(1-y)^3(1-6x-3y+12x^2-4x^3)\ln x \bigr] - [x \leftrightarrow y]\bigr \},\\[1ex]\nonumber
F_{\rm Vr}(x,y) &= \tfrac{2y}{(1-x)^3(1-y)^3(x-y)^3} \bigl \{\bigl [
2(1-x)(1-y)(x^3(4-5y+4y^2) - x^2(2+3y) + x)\\\nonumber
&\hspace*{2cm}+(1-y)^3(4x^4+4x^3y+x^2(3-12y)+x(3y-1)-y)\ln x \bigr] - [x \leftrightarrow y]\bigr \},
\end{align}
where $x = M^2_{\chi}/M_{\eta}^2$ and $y = M_{\chi}^2/\Lambda_c^2$ are the ratios of the DM mass $M_{\chi}$ with respect to the mediator mass $M_{\eta}$ and compositeness scale $\Lambda_c$, respectively.

In Ref.~\cite{lep2} the Wilson coefficients are parametrized in terms of an effective scale $\Lambda_A$ via $\mathcal{C}_{A} = \pm 4\pi/\Lambda_A^2$.
For the comparison with mono-photon constraints from Ref.~\cite{Fox:2011fx}, we translate the LEP limits from $e^+e^- \to \mu^+\mu^-$ interactions for the effective operators $\mathcal{O}_{RR}$ and $\mathcal{O}_{LL}$ to the $90\%$ C.L.,
\begin{equation}\label{eq:Wilson}
\vert\mathcal{C}_{RR}\vert = \frac{4\pi}{\Lambda_{RR}^2} < \frac{4\pi}{(10.2\, (12.7) \tev)^2}, \qquad
\vert\mathcal{C}_{LL}\vert = \frac{4\pi}{\Lambda_{LL}^2} < \frac{4\pi}{(10.7\, (13.3) \tev)^2},
\end{equation}
for $\mathcal{C}_{A}>0$ ($\mathcal{C}_{A}<0$). For the $t$-channel models in Tab.~\ref{tab:mod} they translate into the following bounds:

\begin{figure}[!tb]
\centering
\includegraphics[height=2.4in]{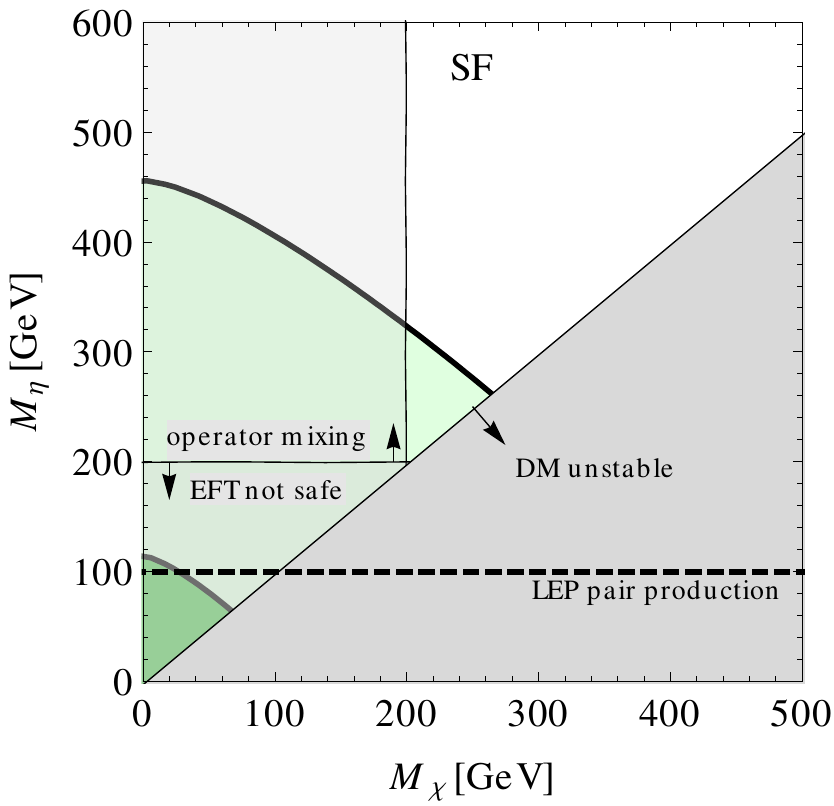}\hspace*{.5cm}
\includegraphics[height=2.4in]{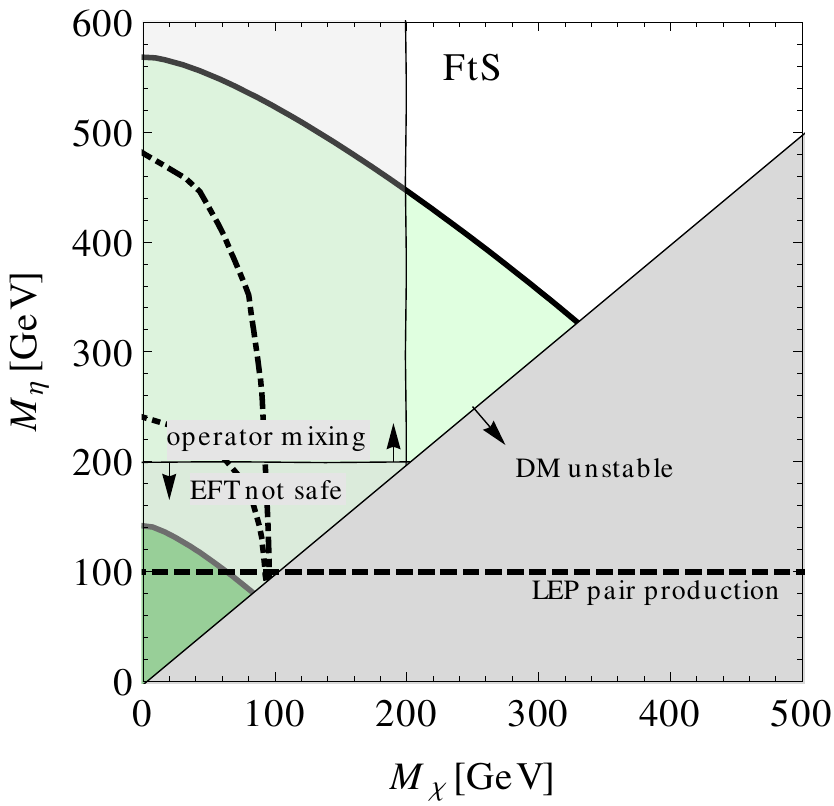}\\[0.5cm]
\includegraphics[height=2.4in]{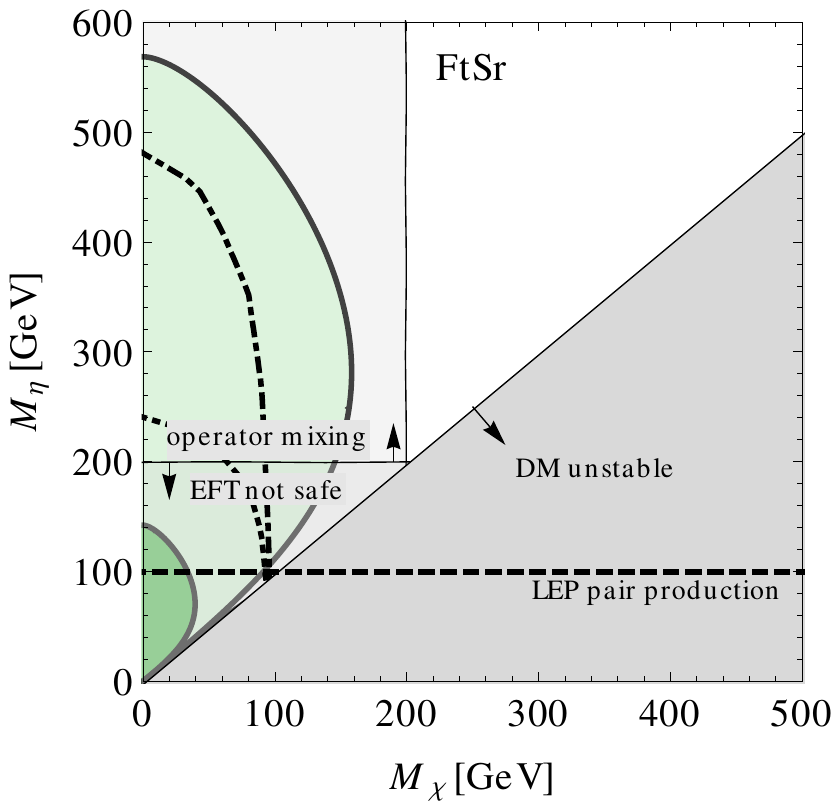}
\vspace{-1ex}
\caption{Four-lepton LEP limits on scenarios with scalars at 90\% C.L.\ for $g_R=1$ (dark green) and $g_R=2$ (light green). In gray regions, the validity of the EFT interpretation or DM stability are not ensured. The 90\% C.L.\ limits from mono-photon events \cite{Fox:2011fx} are displayed for $g_R=1$ ($g_R=2$) as a dotted (dot-dashed) line.}
\label{fig:FS}
\end{figure}
\paragraph{SF, FtS, FtSr:} The constraints on scenarios with scalars stemming from the diagrams in Fig.~\ref{fig:diags}~(a) (SF), Fig.~\ref{fig:diags}~(c) (FtS), and Figs.~\ref{fig:diags}~(c,d) (FtSr) are shown in Fig.~\ref{fig:FS} for two choices of the coupling strength $g_r=1,2$. 
 The plots are for right-handed couplings, but look very similar for the left-handed case. For the scenarios SF and FtS, the Wilson coefficients are symmetric under $M_{\chi}\leftrightarrow M_{\eta}$ and identical up to a sign. The sign difference results in somewhat stronger constraints for FtS, for which $\mathcal{C}_A < 0$ (see (\ref{eq:Wilson})). 
In general, for a coupling strength of about $g=1$ the limits (displayed in dark green) are rather weak. As mentioned earlier, in this region the EFT description is not reliable (see the gray regions). However, for larger couplings $g\gtrsim 2$, one obtains interesting constraints (marked in light green), which extend to masses far beyond the kinematical limit for direct production, $M_\chi > \sqrt{s}/2 \sim 100\gev$. 

For comparison, we show limits on direct DM production from mono-photon searches $e^+ e^- \rightarrow \chi\chi\gamma$ at LEP \cite{Fox:2011fx} in the scenarios FtS and FtSr for chiral couplings $g_r=1$ (dotted line) and $g_r=2$ (dot-dashed line)\footnote{To account for chiral couplings, the limits from Ref.~\cite{Fox:2011fx} were divided by a factor of $\sqrt{2}$.}. For moderate couplings $g\lesssim 1$, the sensitivity to DM effects is comparable with four-lepton interactions. For larger couplings, four-lepton interactions clearly yield stronger bounds than mono-photon searches, which are always confined to $M_{\chi} < 100\gev$. For the other $t$-channel scenarios in Tab.~\ref{tab:mod}, mono-photon limits have not been obtained in Ref.~\cite{Fox:2011fx}, but are expected to be of similar strength.

\paragraph{SFr:} For the case of real scalar DM, the contributions of the two diagrams in Figs.~\ref{fig:diags}~(a) and (b) cancel exactly in the EFT limit. Therefore no bound is obtained from LEP four-lepton contact interactions. In this case, mono-photon searches are expected to provide the strongest constraints from LEP.

\begin{figure}[tb]
\centering
\begin{tabular}{cc}
\includegraphics[height=2.4in]{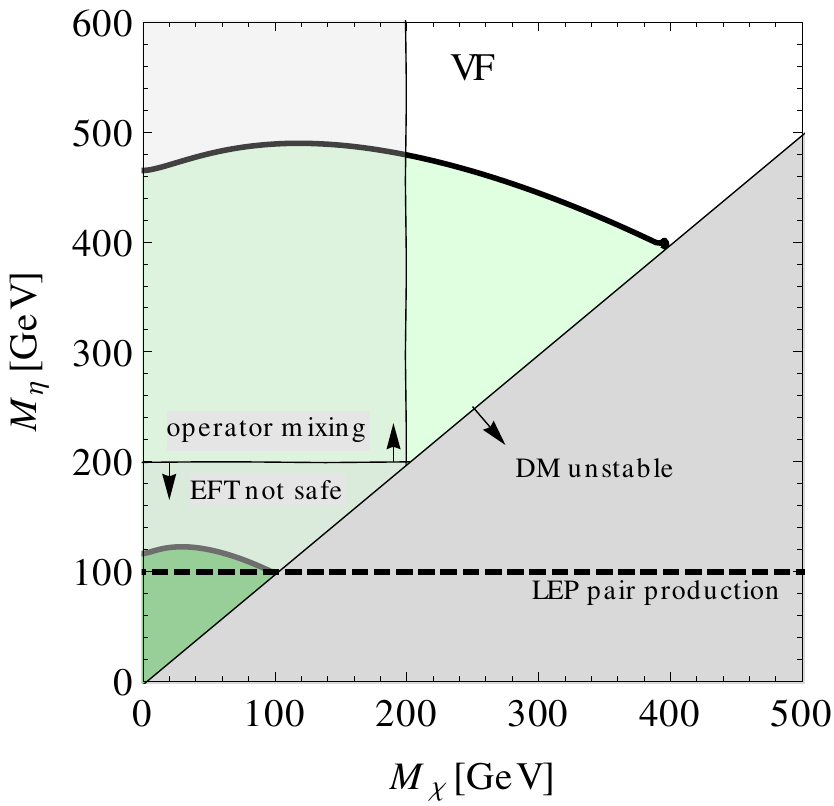} &
\includegraphics[height=2.4in]{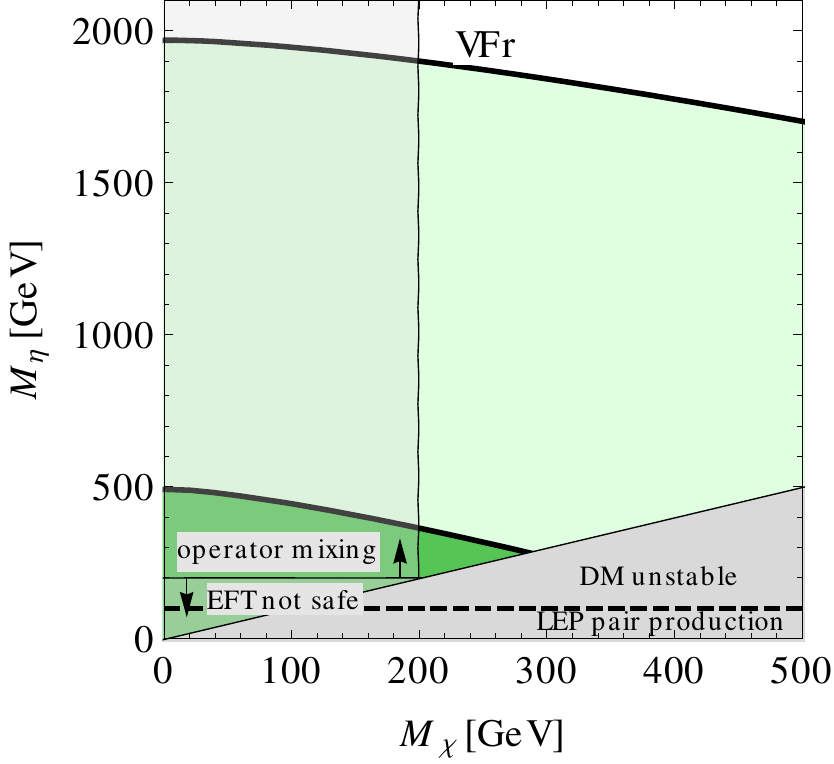}\\[0.5cm]
\includegraphics[height=2.4in]{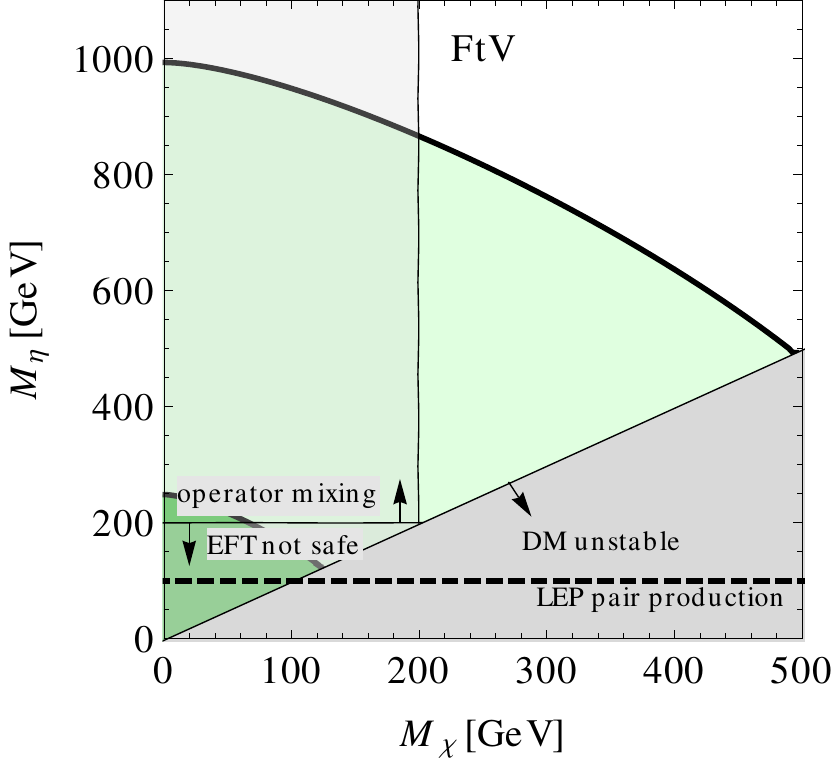} &
\includegraphics[height=2.4in]{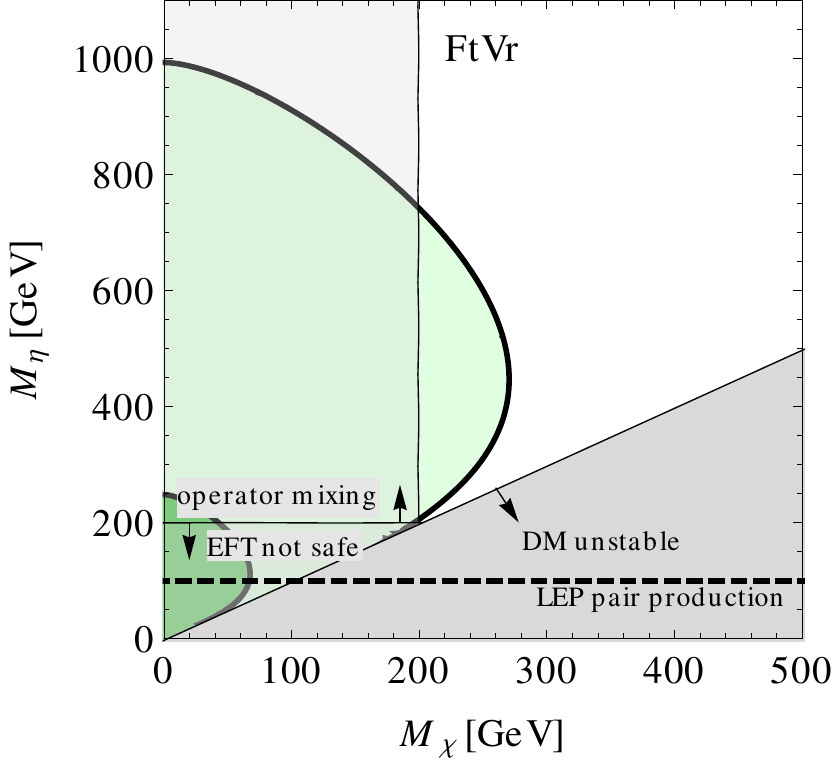}
\end{tabular}
\caption{Four-lepton LEP limits on scenarios with vectors at 90\% C.L. for $g_R=1$ (dark green) and $g_R=2$ (light green). In the scenarios VF, FtV, and FtVr, the vector bosons are assumed to be composite with an associated scale $\Lambda_c = M_V$. In gray regions, the EFT interpretation is not safe or DM is unstable.}
\label{fig:FV}
\end{figure}

\paragraph{VFr:} Contributions to four-lepton interactions from real vector-boson DM with a fermion mediator originate from the diagrams in Figs.~\ref{fig:diags}~(e) and (f). As has been discussed above (\ref{eq:prop}), the sum of both diagrams is gauge-independent and finite. For chiral couplings, the mass dependence of the result happens to be the same as in the scenario FtS, but the Wilson coefficient is enhanced by a factor of 12 (see Tab.~\ref{tab:4lci}). The corresponding constraints on VFr are shown for $g_r=1$ (dark green) and $g_r=2$ (light green) in the upper right panel of Fig.~\ref{fig:FV}. The bounds extend well beyond the kinematic limit of direct DM production at LEP already for moderate couplings $g\approx 1$.

\begin{figure}[tbp]
\centering
\includegraphics[scale=1]{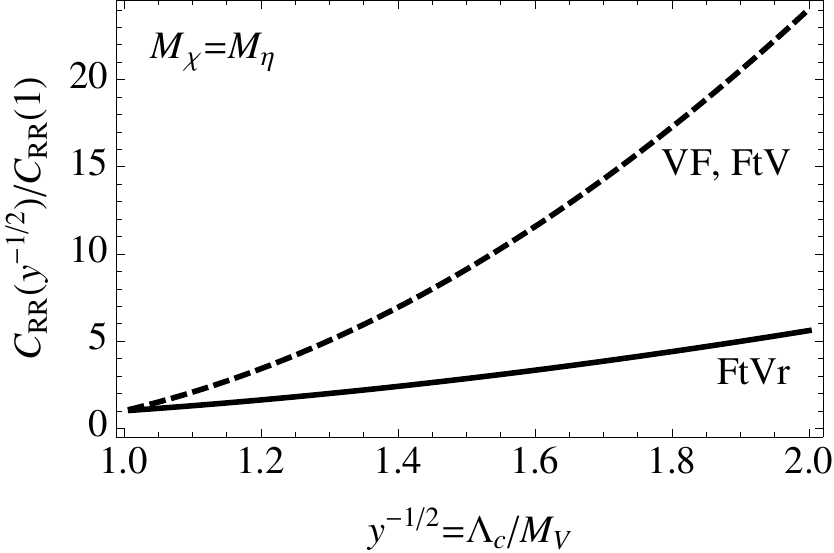}
\caption{Wilson coefficient $\mathcal{C}_{RR}$ in scenarios with composite vector bosons. Shown is the dependence on the scale $\Lambda_c$ for the scenarios FtVr (plain) and VF, FtV (dashed). DM and mediator masses are set equal, $M_{\chi} = M_{\eta}$.}
\label{fig:composite}
\end{figure}

\paragraph{VF, FtV, FtVr:} In the presence of (composite) complex vector bosons, four-lepton interactions are induced by the diagrams in Fig.~\ref{fig:diags}~(e) (VF), Fig.~\ref{fig:diags}~(g) (FtV, FtVr), and Figs.~\ref{fig:diags}~(h) (FtVr). The corresponding LEP bounds are shown in Fig.~\ref{fig:FV} for $g_r=1,2$ and $\Lambda_c=M_V$. As can be observed from Tab.~\ref{tab:4lci}, the Wilson coefficients $\mathcal{C}_{RR}$ in all three scenarios exhibit a quadratic dependence on the compositeness scale $\Lambda_c$. We show this feature explicitly in Fig.~\ref{fig:composite} for equal DM and mediator masses.
 The strong dependence on $\Lambda_c$ suggests that the compositeness scale should not lie far above the mass of the vector boson $M_V$, so that
 the damping of the propagator by the form factor in (\ref{eq:form}) is effective for momenta $k^2 \gtrsim M_V^2$. The constraints from Fig.~\ref{fig:FV} with $\Lambda_c=M_V$ thus provide a conservative estimate of DM effects in composite scenarios. Precise bounds in a specific model with composite vector bosons will significantly depend on the realization of strong dynamics at the scale $\Lambda_c$.


\section{Projections for the ILC}
\label{ilc}
At a future linear $e^+ e^-$ collider with a CM energy up to $1\tev$, such as ILC, the sensitivity to leptophilic DM scenarios is expected to increase significantly with respect to LEP. In this section, the reach of the
ILC will be estimated by re-scaling the LEP limits by the following factors.
\begin{itemize}
\item The new-physics amplitude $\mathcal{A}_{\rm NP}$ induced by the effective operators \eqref{eq:ops} increases with the CM energy $\sqrt{s}$ relative to the amplitude for the SM background, 
\mbox{$|{\cal A}_{\rm NP}/{\cal A}_{\rm SM}| \propto s/\Lambda^2$}. Assuming Poisson statistics, the statistical uncertainty on the Wilson coefficients $\mathcal{C}_A$ therefore scales $\propto s^{-1/2}$ for higher energies.
\item The increased luminosity $\cal L$ leads to an enhancement of the signal and background cross-sections and thus to a reduction of the statistical error $\propto {\cal L}^{-1/2}$.
\item The signal yield can be further enhanced with the help of beam polarization. For new physics in the $O_{LL}$ ($O_{RR}$) operator, the optimal choice are left-handed (right-handed) incoming electrons and right-handed (left-handed) incoming positrons. If the electron/positron polarization degree is denoted by $P^-/P^+$, the signal rate is enhanced by the factor $r_S = N_S/N_S^{\rm unpol}=(1+P^-)(1+P^+)$ compared to the unpolarized case.\footnote{Here $N_S$ ($N_B$) and $N_S^{\rm unpol}$ ($N_B^{\rm unpol}$) denote the total number of polarized and unpolarized signal (background) events, respectively.}
Similarly, for scalar and vector operators, the signal with optimal polarization is enhanced by $r_S = (1+P^-P^+)$.
\item Polarization also changes the statistical uncertainty of the SM background $e^+e^-\rightarrow \gamma/Z\rightarrow \mu^+\mu^-$ by $r_B^{1/2}=(N_B/N_B^{\rm unpol})^{1/2}$. A background enhancement due to polarization thus reduces the statistical significance of the signal by $r_B^{-1/2}$.
\end{itemize}
In summary, the expected reduction of the statistical uncertainty for constraining four-fermion contact interactions at the ILC as compared to LEP leads to improved limits on the Wilson coefficients, given by
\begin{align}\label{eq:ilc-limits}
|\mathcal{C}_{\rm LL,RR}|^{\rm max}_{\rm ILC} 
= |\mathcal{C}_{\rm LL,RR}|^{\rm max}_{\rm LEP} \times \biggl [\frac{s_{\rm ILC}}{s_{\rm LEP}} \times \frac{{\cal L}_{\rm ILC}}{{\cal L}_{\rm LEP}} \biggr ]^{-1/2} \times \frac{\sqrt{r_B}}{r_S},
\end{align}
where $r_{S,B}$ denote the signal and background enhancement due to beam polarization, respectively, as defined in the items above. We consider the high-energy option of the ILC with $\sqrt{s}_{\rm ILC}=1\tev$ and a luminosity of ${\cal L}_{\rm ILC}=500\text{ fb}^{-1}$. We further take $\sqrt{s}_{\rm LEP}\approx 200\gev$ as the approximate CM energy where most of the data were accumulated at LEP2 and the combined luminosity of four LEP experiments, ${\cal L}_{\rm LEP}=4\times 0.75\text{ fb}^{-1}$. We choose the polarizations $P^-=0.8$ and $P^+=0.6$. A polarization degree of 60\% for positrons is very optimistic, but we will use this value in our numerical analysis, so that we can compare with the mono-photon results from Ref.~\cite{Dreiner:2012xm}. Using the program {\sc CalcHEP} \cite{Belyaev:2012qa}, we have estimated the background error variation due to beam polarization. For an $e^+_Re^-_L$ ($e^+_Le^-_R$) beam polarization, we find $\sqrt{r_B}\approx 1.3\ (1.2)$. Our numerical analysis will be performed in the scenario $e^+_Le^-_R$ with dominantly right-handed electrons and left-handed positrons, which leads to the maximal statistical significance for the vector and right-handed benchmark scenarios in four-lepton interactions and mono-photon searches (see Tab.~VIII in Ref.~\cite{Dreiner:2012xm}).

It is expected that the bulk of the experimental systematics for the measurement of $e^+e^- \to \ell^+\ell^-$ will go down in parallel with the statistical error, since the calibration and determination of the experimental efficiency is improved with more statistics. Systematic uncertainties in the measurement of the polarization degree can be neglected, since they contribute only at the percent level to the total background error, provided the polarization at ILC is known to a precision of about $0.1\%$. Furthermore, we will assume that the theoretical uncertainties will be reduced by future higher-order calculations, such that they do not limit the precision of the four-fermion contact interaction bounds. In our numerical analysis, systematic uncertainties will thus be treated by simply scaling the LEP limits with the factor \eqref{eq:ilc-limits}.

\paragraph{S-channel scenarios:} The LEP limits on four-lepton contact interactions through $s$-channel mediators from (\ref{eq:s-v}) and (\ref{eq:s-s}) are projected onto the polarized $e^+_L e^-_R$ ILC setup using (\ref{eq:ilc-limits}).
For vector mediators, we obtain the following ILC bounds at the $90\%$ C.L.:
\begin{align}
&\text{vector:} & g_v/M_\eta &< 2.2\times 10^{-5} \gev^{-1} & &(M_\eta>1\tev), \\
& & g_v/M_\eta &< 7.6\times 10^{-5} \gev^{-1} & &(100\gev<M_\eta<1\tev), \\[1ex]
&\text{axial-vector:} & g_a/M_\eta &< 2.7\times 10^{-5} \gev^{-1} & &(M_\eta>1\tev), \\
& & g_a/M_\eta &< 7.6\times 10^{-5} \gev^{-1} & &(100\gev<M_\eta<1\tev).
\end{align}
For scenarios with scalar mediators, the production rate for $e^+_Le^-_R$ beam polarization is suppressed by a factor of $(1-P^-P^+)$. We therefore give the projected limits for an unpolarized setup. Using (\ref{eq:ilc-limits}) with $\sqrt{r_B}/r_S=1$, we obtain
\begin{align}
&\text{(pseudo)scalar:} & g_{s,p}/M_\eta &< 3.4\times 10^{-5} \gev^{-1} & &(M_\eta>1\tev), \\ & & g_{s,p}/M_\eta &< 9.1\times 10^{-5} \gev^{-1} & &(100\gev<M_\eta<1\tev).
\end{align}
For mediator masses below $200\gev$, ILC bounds are thus expected to be about one order of magnitude stronger than LEP limits. For larger masses, the ILC has the potential to test scenarios with heavy mediators beyond $10\tev$, if couplings are of $\mathcal{O}(1)$.

\begin{figure}[tb]
\centering
\begin{tabular}{cc}
\includegraphics[height=2in]{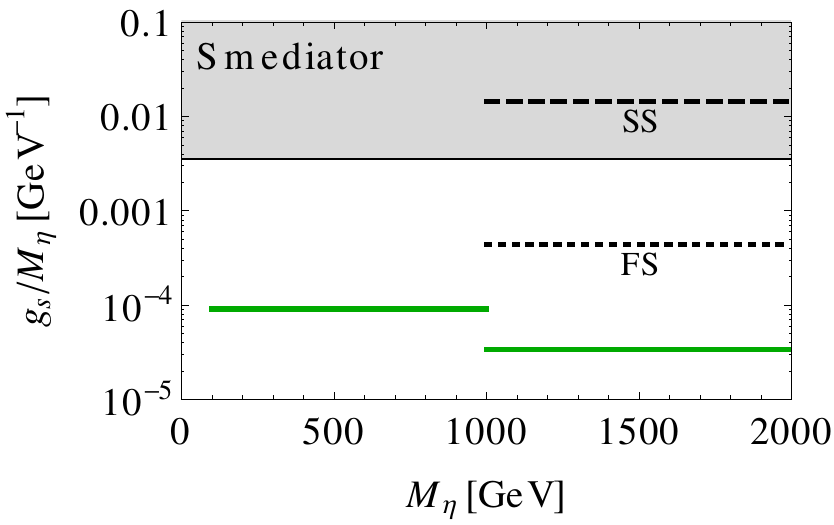} &
\includegraphics[height=2in]{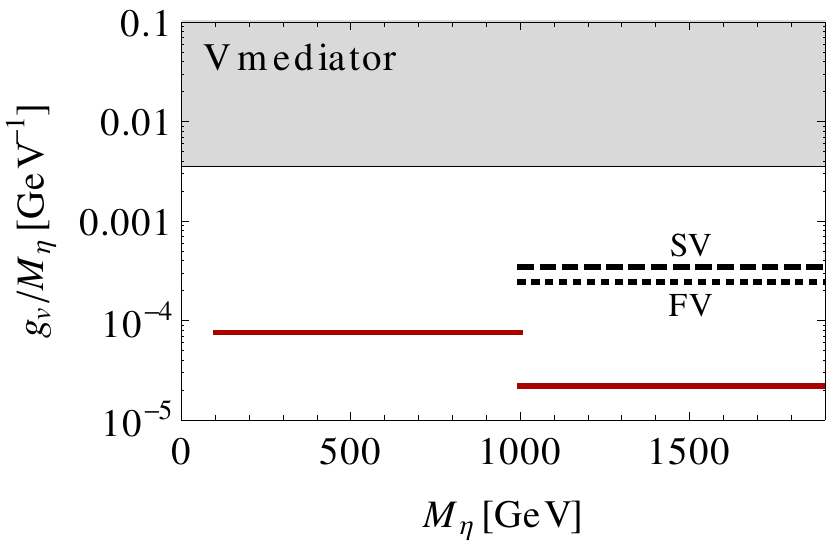}
\end{tabular}
\caption{ILC sensitivity to $s$-channel DM scenarios with scalar (left) and vector (right) mediators for $\sqrt{s}=1\tev$ in an unpolarized (left) and polarized (right) setup. Shown are projected $90\%$ C.L. upper bounds on the effective coupling $g/M_{\eta}$ from four-lepton interactions (plain green/red) and mono-photon searches \cite{Dreiner:2012xm} (SS, SV: dashed black; FS, FV: dotted black). Mono-photon bounds get weaker for $M_{\chi} \gtrsim 200\gev$ and vanish at the pair production threshold $M_{\chi} = 500\gev$. In the gray region, an EFT description with $g < \sqrt{4\pi}$ is not reliable.}
\label{fig:ilc-s}
\end{figure}

In Fig.~\ref{fig:ilc-s}, we compare the ILC sensitivity to scenarios with scalar (left panel) or vector (right panel) mediators for four-lepton interactions (green/red lines) and mono-photon searches (black dashed for scalar DM, black dotted for fermion DM).\footnote{We do not show mono-photon bounds on scenarios with vector DM, since they depend strongly on the UV completion of the respective model, see the discussion in Sec.~\ref{loop}.} The estimates for mono-photon searches are taken from Fig.~6 in Ref.~\cite{Dreiner:2012xm}, where the bounds on scalar mediators have been rescaled to an unpolarized ILC setup by $(g_S/M_\eta)^{\rm unpol}=(g_S/M_\eta)\times(r_S/\sqrt{r_B})^{1/4}$. Unlike four-lepton limits, mono-photon limits are restricted to $M_{\eta} > \sqrt{s}=1\tev$, where the description in terms of effective $ee\chi\chi$ couplings is valid. In this high-mass regime, the bounds from four-lepton interactions exceed the mono-photon bounds by one order of magnitude.

\begin{figure}[!tb]
\centering
\includegraphics[height=2.4in]{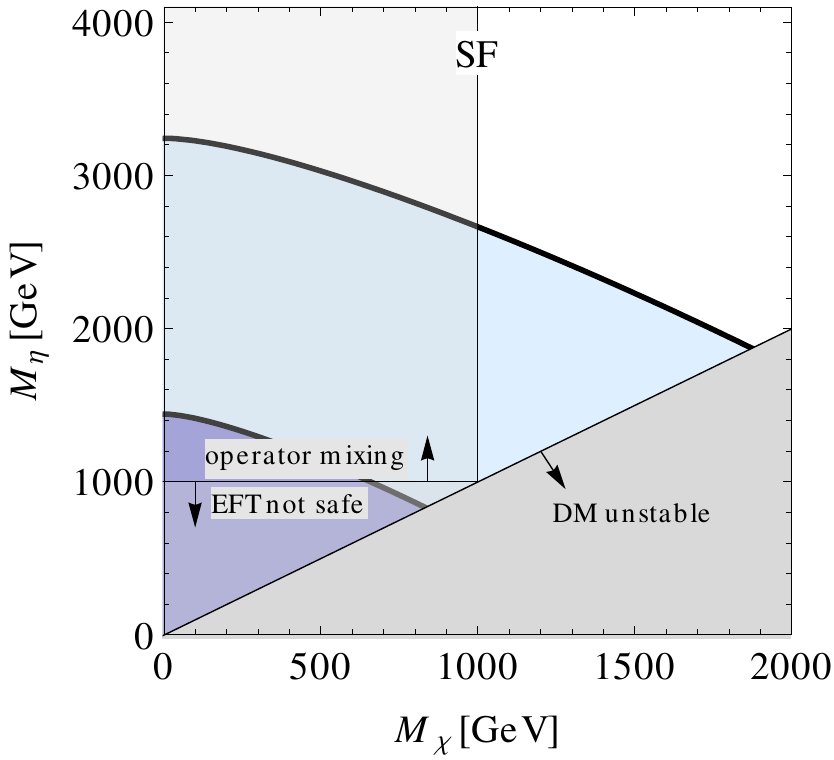}\hspace*{.5cm}
\includegraphics[height=2.4in]{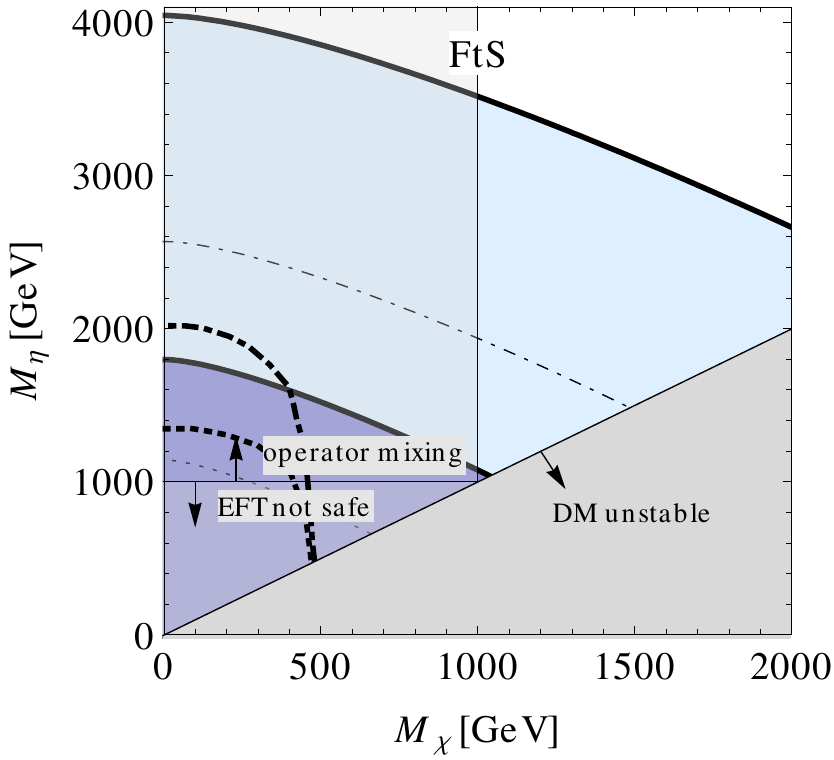}\\[0.5cm]
\includegraphics[height=2.4in]{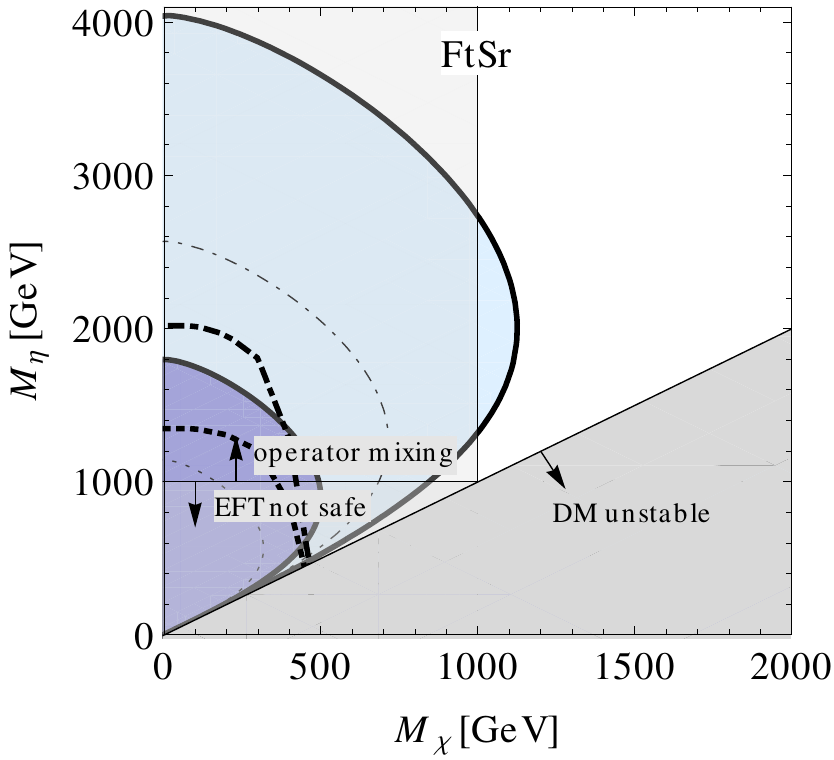}
\vspace{-1ex}
\caption{ILC sensitivity to $t$-channel DM scenarios with scalars for $\sqrt{s}=1\tev$. Shown are projected 90\% C.L. regions from four-lepton interactions for $g_R=1$/$g_R=1.5$ in a polarized setup (dark blue/light blue) and an unpolarized setup (thin dotted/thin dot-dashed line). Bounds from mono-photon searches \cite{Dreiner:2012xm} are displayed in an unpolarized setup (thick dotted/thick dot-dashed line).}
\label{fig:t-scalar-ilc}
\end{figure}

\begin{figure}[!tb]
\centering
\begin{tabular}{cc}
\includegraphics[height=2.4in]{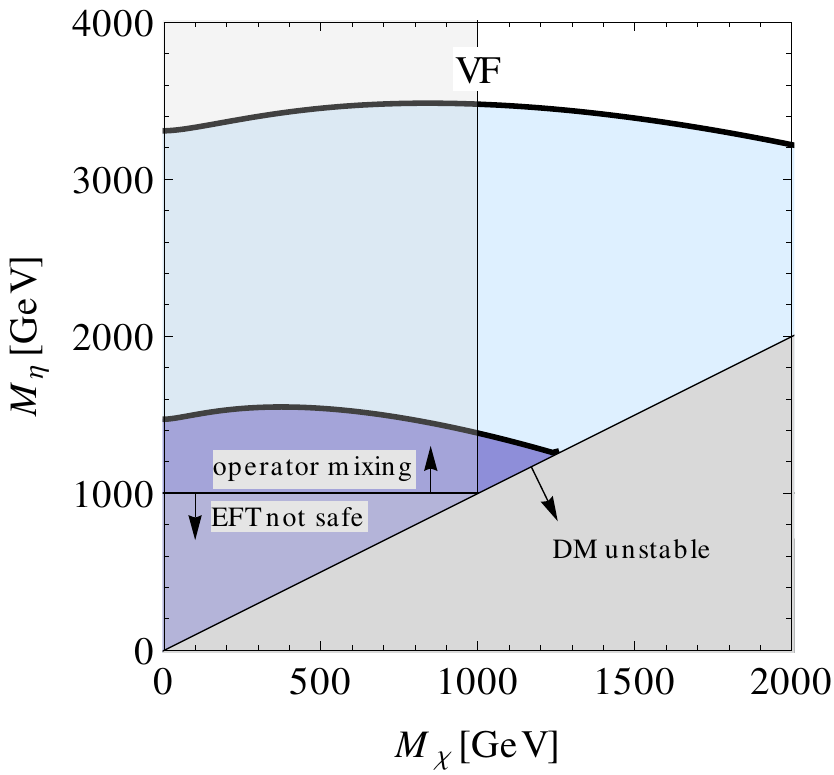} &
\includegraphics[height=2.4in]{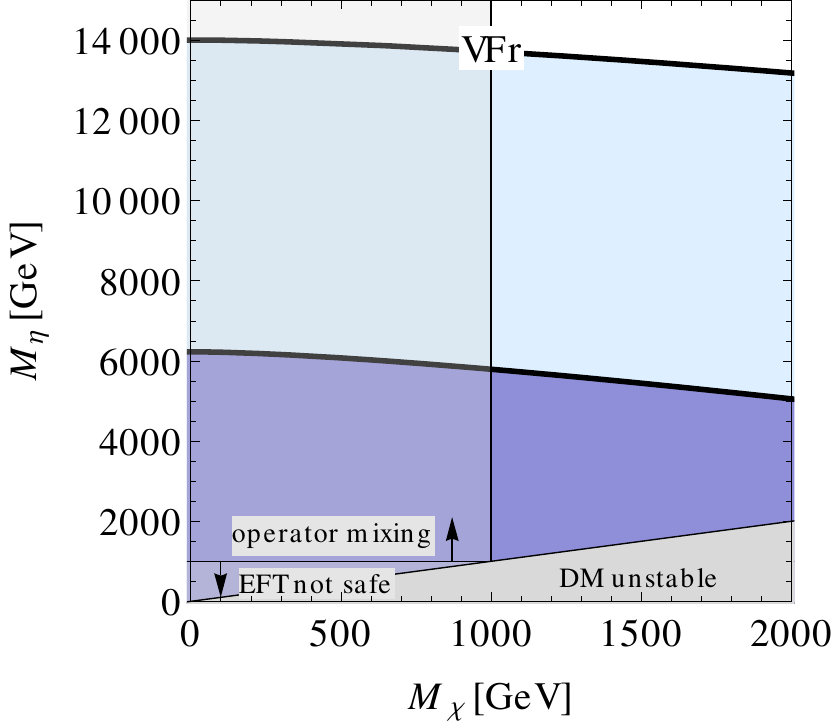}\\[0.5cm]
\includegraphics[height=2.4in]{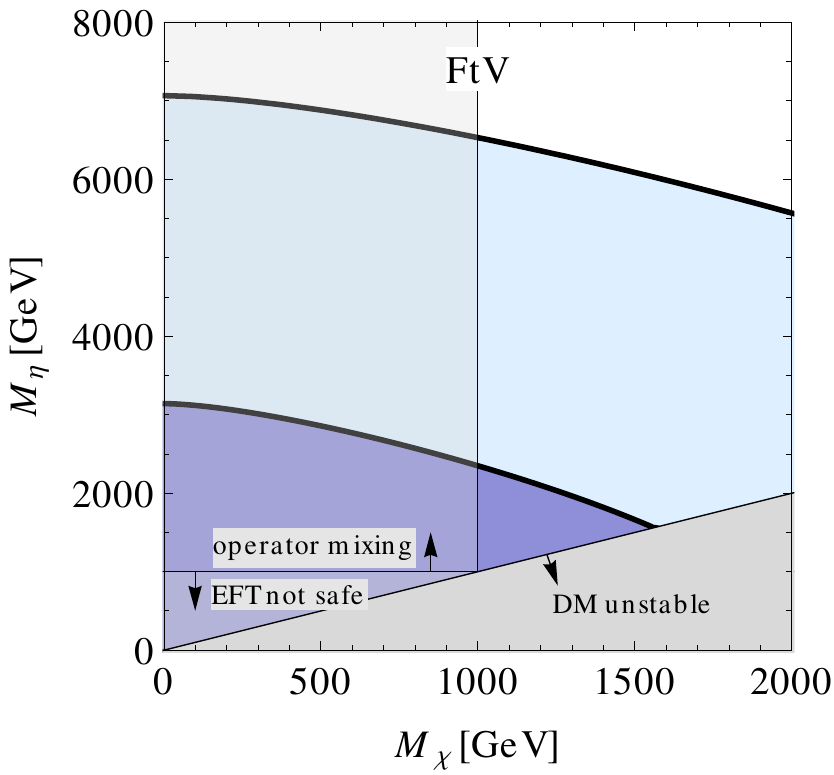} &
\includegraphics[height=2.4in]{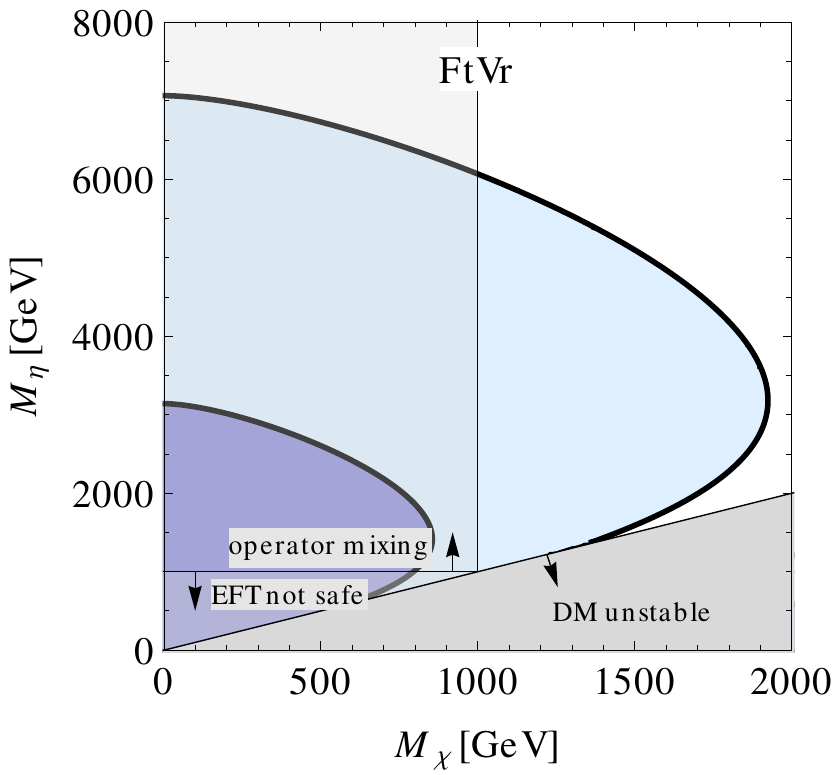}
\end{tabular}
\caption{ILC sensitivity to $t$-channel DM scenarios with vectors for $\sqrt{s}=1\tev$ and polarizations $P^-=0.8$, $P^+=0.6$. Shown are projected 90\% C.L. regions from four-lepton interactions for $g_R=1$ (dark blue) and $g_R=1.5$ (light blue). In the scenarios VF, FtV, and FtVr, the vector bosons are assumed to be composite with an associated scale $\Lambda_c = M_V$.}
\label{fig:t-vector-ilc}
\end{figure}

\paragraph{T-channel scenarios:} We estimate the ILC reach for scenarios with $t$-channel mediators (Tab.~\ref{tab:4lci}) by rescaling LEP limits on four-lepton interactions \cite{lep2} using the projection from (\ref{eq:ilc-limits}). The corresponding $90\%$ C.L.\ limits on the parameter space are shown in Fig.~\ref{fig:t-scalar-ilc} for scenarios with scalars and in Fig.~\ref{fig:t-vector-ilc} for scenarios with vector bosons for the couplings $g_r=1$ (dark blue) and $g_r=1.5$ (light blue). Our polarization choice $e^+_Le^-_R$ is particularly suitable for $t$-channel scenarios with right-chiral couplings, as all of them are described by one single effective operator $\mathcal{O}_{RR}$. It is immediately apparent that four-lepton interactions at the ILC will be sensitive to DM scenarios with particles in the multi-TeV range. The lower-mass region is constrained by LEP results (see Figs.~\ref{fig:FS} and \ref{fig:FV}). The sensitivity to scenarios with left-chiral couplings ($\mathcal{O}_{LL}$) is very similar, assuming an optimal polarization $e^+_Re^-_L$. A bound on the Wilson coefficient $\mathcal{C}_{LL}$ is obtained from the right-chiral results by correcting for the different LEP limits (see (\ref{eq:Wilson})) and background uncertainty $\sqrt{r_B}$, yielding $|\mathcal{C}_{LL}|^{\rm max}_{\rm ILC}=|\mathcal{C}_{RR}|^{\rm max}_{\rm ILC}\times(|\mathcal{C}_{LL}|^{\rm max}_{\rm LEP}/|\mathcal{C}_{RR}|^{\rm max}_{\rm LEP}) \times (1.3/1.2)$.

In the scenarios with scalars, we compare our results from four-lepton interactions to the mono-photon searches from Ref.~\cite{Dreiner:2012xm}.$^3$ In order to allow a direct comparison, the bounds from both observables have been rescaled to an unpolarized ILC setup. The results are displayed in Fig.~\ref{fig:t-scalar-ilc} in the scenarios FtS and FtSr with fermion DM for $g_r=1$ (dotted lines) and $g_r=1.5$ (dot-dashed lines). For $g_r=1$ the sensitivity is comparable, while for larger couplings four-lepton interactions extend to much larger masses than mono-photon events. In the scenarios SF and SFr, the leading contribution to mono-photon signals vanishes for chiral couplings. The SF scenario can thus be tested only with four-lepton interactions. The scenario SFr cannot be probed by four-lepton interactions (which are absent for chiral couplings), and will be very difficult to access with mono-photon searches.



\section{Discussion}
\label{disc}

In Ref.~\cite{Dreiner:2012xm} the expected reach from mono-photon searches at ILC
was compared to limits from direct and indirect detection. For models where the dark sector couples flavor-universally to leptons, but not to quarks, the constraints from direct detection are loop-suppressed. As a result, the projected mono-photon bounds are stronger than the present-day direct detection limits for kinematically accessible DM masses at the ILC ($M_\chi \lesssim 500\gev$). For the same mass range, they are also stronger than indirect detection constraints from the PAMELA and AMS-02 experiments.

As shown in the previous section, for $s$-channel mediated models the projected bounds from four-lepton contact interactions significantly exceed the reach of mono-photon searches at ILC. Therefore they will be the strongest constraints on this class of models. Moreover, they are independent of $M_\chi$ and thus can cover the entire parameter space that is probed by direct detection experiments. 

For $t$-channel mediated models, the four-lepton bounds are comparable to mono-photon bounds for $M_\chi \lesssim 500\gev$ and couplings $g \gtrsim {\cal O}(1)$. However, the four-lepton limits extend also to larger DM masses, where they will be competitive with current and future direct detection limits.

\vspace{\medskipamount}
If the DM and mediator fields are part of large electroweak SU(2) gauge multiplets, the additional states in these multiplets will also contribute to the four-lepton contact interactions. For small SU(2) representations, such as doublets or triplets, the sum of extra diagrams enhances the expected new-physics signal with respect to the minimal contributions studied in this work. Therefore, our results can still be used as conservative bounds for simple non-minimal dark sectors. However, more complex dark sectors, for example involving new types of interactions, may lead to destructive interference in the $ee\mu\mu$ operators, so that our results are not directly applicable to these cases.

\vspace{\medskipamount}
Leptophilic DM models are additionally motivated by their ability to explain the observed discrepancy $\Delta a_\mu$ between the measured value and the SM prediction for the muon anomalous magnetic moment \cite{Freitas:2014pua,Agrawal:2014ufa}. For models with $t$-channel mediators, the correct value for $\Delta a_\mu$  can be obtained for dark-sector masses $M_{\chi,\eta}$ of a few hundred GeV. This parameter region is also probed by four-fermion contact interactions. In fact, the LEP limits already exclude the majority of the preferred parameter space for $\Delta a_\mu$, except for self-conjugate scalar DM, where the four-lepton contributions vanish \cite{Freitas:2014pua}.

\vspace{\medskipamount}
The results in this paper have been obtained by adopting the EFT framework used by the LEP collaborations. The EFT is formally only valid for $M_\eta > \sqrt{s}$. Furthermore, for $M_\chi < \sqrt{s}$ additional operators may need to be included for a fully consistent description, although their numerical impact is expected to be subdominant, as discussed in section~\ref{loop}. As evident from Figs.~\ref{fig:FS}, \ref{fig:FV}, \ref{fig:t-scalar-ilc} and \ref{fig:t-vector-ilc}, these conditions are only fulfilled for models with $t$-channel mediator if the $\ell$-$\chi$-$\eta$ coupling is larger than about one. For smaller couplings, the EFT treatment breaks down. Instead one would need to compare exact results for the relevant box diagrams with the experimental data for the process $e^+e^- \to \ell^+\ell^-$. Such an analysis is beyond the scope of this work. But our results provide an approximate indication of the achievable limits in these weak-coupling scenarios.


\section{Conclusions}
\label{concl}

In this work, we have investigated the sensitivity of $e^+ e^-\rightarrow \ell^+\ell^-$ interactions to minimal models of leptophilic DM. We have considered scenarios with $s$- and $t$-channel mediation in an EFT, where the mediator and DM fields do not appear as dynamical degrees of freedom. The low-energy imprints of the dark sector are encoded in effective four-lepton interactions, which have been strongly constrained by LEP and can be tested at a future $e^+ e^-$ collider.

In models with $s$-channel mediation, four-lepton interactions are sensitive to the mediator field and its coupling to SM leptons only. LEP bounds on these interactions limit the coupling-to-mass ratio to $g/M_{\eta} < \mathcal{O}(10^{-4})\gev^{-1}$ at the $90\%$ C.L. $S$-channel mediators with masses below the terascale are thus excluded by LEP, if their coupling to leptons is of $\mathcal{O}(1)$.

Scenarios with $t$-channel mediation induce four-lepton interactions at the one-loop level, involving both the mediator and the DM fields. We have focused on chiral interactions with leptons, where effects on four-lepton interactions can be expressed by one single chiral vector-current operator $\mathcal{O}_{LL}$ or $\mathcal{O}_{RR}$. Due to the loop suppression, LEP bounds on $t$-channel mediated scenarios are rather weak for couplings $g\lesssim 1$. For larger couplings, thanks to the $g^4/M^2$ scaling of the effective four-lepton interaction, LEP bounds for scenarios with scalars extend to masses of a few hundred GeV. In scenarios with complex vector bosons, the loop-induced four-lepton interactions are gauge-dependent and thus require an embedding into a complete theory at high energies. We have considered a framework with composite vector bosons and implemented an energy cutoff by introducing a simple form factor. The resulting bounds depend strongly on the compositeness scale $\Lambda_c$. With the conservative assumption that $\Lambda_c$ lies close above the vector-boson mass, LEP bounds on scenarios with composite vectors are slightly stronger than for scalars.

At a future ILC, the sensitivity to leptophilic DM can be significantly enhanced due to three main factors: a larger CM energy, higher luminosity and the possibility of polarized beams. We have estimated the reach of ILC by rescaling the LEP bounds with respect to these three factors. As a result, the ILC sensitivity to $s$-channel mediators increases by one order of magnitude with respect to LEP, yielding $g/M_{\eta} < \mathcal{O}(10^{-5})\gev^{-1}$. Similarly, $t$-channel scenarios can be probed up to masses in the multi-TeV range for ${\cal O}$(1--2) couplings. 

It is interesting to compare these results with mono-photon signals $e^+ e^-\rightarrow \chi\chi\gamma$, hitherto considered one of the most efficient observables to constrain DM at colliders. In weak-coupling scenarios, the sensitivity of four-lepton interactions is comparable to mono-photon signals at LEP and ILC. However, for couplings of strength larger than one, four-lepton interactions cover a DM mass range way above the production limit in mono-photon searches, $M_{\chi} < \sqrt{s}/2$.

In the future, precision measurements of the 
process $e^+ e^-\rightarrow \ell^+\ell^-$ will yield information on leptophilic DM models that goes beyond effective four-lepton interactions. Differential distributions and polarization-dependent observables at the ILC can probe the structure of the dark sector in much more detail. Virtual effects in lepton interactions are thus an interesting alternative and an important complement to mono-photon searches in exploring leptophilic DM at colliders.


\subsection*{Acknowledgments}
This work was supported in part by the National Science Foundation, grant PHY-1212635.




\begin{thebibliography}{99}

\bibitem{dd}
  R.~Agnese {\it et al.}  [SuperCDMS Collaboration],
  Phys.\ Rev.\ Lett.\  {\bf 112}, 041302 (2014)
  [arXiv:1309.3259 [physics.ins-det]];\\
  D.~S.~Akerib {\it et al.}  [LUX Collaboration],
  Phys.\ Rev.\ Lett.\  {\bf 112}, 091303 (2014)
  [arXiv:1310.8214 [astro-ph.CO]];\\
  R.~Agnese {\it et al.}  [SuperCDMS Collaboration],
  Phys.\ Rev.\ Lett.\  {\bf 112}, 241302 (2014)
  [arXiv:1402.7137 [hep-ex]].

\bibitem{id} 
  M.~Ackermann {\it et al.}  [Fermi-LAT Collaboration],
  Phys.\ Rev.\ D {\bf 89}, 042001 (2014)
  [arXiv:1310.0828 [astro-ph.HE]];\\
  M.~Cirelli and G.~Giesen,
  JCAP {\bf 1304}, 015 (2013)
  [arXiv:1301.7079 [hep-ph]].

\bibitem{lhcd}
  G.~Aad {\it et al.}  [ATLAS Collaboration],
  JHEP {\bf 1304}, 075 (2013)
  [arXiv:1210.4491 [hep-ex]]; 
  S.~Chatrchyan {\it et al.} [CMS Collaboration],
  CMS-PAS-EXO-12-048.

\bibitem{leptophil1}
  P.~J.~Fox and E.~Poppitz,
  Phys.\ Rev.\ D {\bf 79}, 083528 (2009)
  [arXiv:0811.0399 [hep-ph]];
J.~Kopp, V.~Niro, T.~Schwetz and J.~Zupan,
  Phys.\ Rev.\ D {\bf 80}, 083502 (2009)
  [arXiv:0907.3159 [hep-ph]].
  
\bibitem{leptophil2}
  R.~Harnik and G.~D.~Kribs,
  Phys.\ Rev.\ D {\bf 79}, 095007 (2009)
  [arXiv:0810.5557 [hep-ph]];\\
  A.~Dedes, I.~Giomataris, K.~Suxho and J.~D.~Vergados,
  Nucl.\ Phys.\ B {\bf 826}, 148 (2010)
  [arXiv:0907.0758 [hep-ph]].

\bibitem{Dreiner:2012xm} 
  H.~Dreiner, M.~Huck, M.~Kr\"amer, D.~Schmeier and J.~Tattersall,
  Phys.\ Rev.\ D {\bf 87}, no. 7, 075015 (2013)
  [arXiv:1211.2254 [hep-ph]].

\bibitem{Fox:2011fx} 
  P.~J.~Fox, R.~Harnik, J.~Kopp and Y.~Tsai,
  Phys.\ Rev.\ D {\bf 84}, 014028 (2011)
  [arXiv:1103.0240 [hep-ph]].

\bibitem{ilc2}
  A.~Birkedal, K.~Matchev and M.~Perelstein,
  Phys.\ Rev.\ D {\bf 70}, 077701 (2004)
  [hep-ph/0403004];\\
  P.~Konar, K.~Kong, K.~T.~Matchev and M.~Perelstein,
  New J.\ Phys.\  {\bf 11}, 105004 (2009)
  [arXiv:0902.2000 [hep-ph]];\\
  C.~Bartels, M.~Berggren and J.~List,
  Eur.\ Phys.\ J.\ C {\bf 72}, 2213 (2012)
  [arXiv:1206.6639 [hep-ex]].

\bibitem{lep2}
  S.~Schael {\it et al.}  [ALEPH and DELPHI and L3 and OPAL and LEP Electroweak Collaborations],
  Phys.\ Rept.\  {\bf 532}, 119 (2013)
  [arXiv:1302.3415 [hep-ex]].

\bibitem{feynarts}
  T.~Hahn,
  Comput.\ Phys.\ Commun.\  {\bf 140}, 418 (2001)
  [hep-ph/0012260].

\bibitem{Mertig:1990an} 
  R.~Mertig, M.~Bohm and A.~Denner,
  Comput.\ Phys.\ Commun.\  {\bf 64}, 345 (1991).

\bibitem{Abbiendi:1999wm}
  G.~Abbiendi {\it et al.}  [OPAL Collaboration],
  Eur.\ Phys.\ J.\ C {\bf 13}, 553 (2000)
  [hep-ex/9908008].

\bibitem{Freitas:2014pua} 
  A.~Freitas, J.~Lykken, S.~Kell and S.~Westhoff,
  JHEP {\bf 1405}, 145 (2014)
  [arXiv:1402.7065 [hep-ph]].

\bibitem{Bell:2014tta}
  N.~F.~Bell, Y.~Cai, R.~K.~Leane and A.~D.~Medina,
  arXiv:1407.3001 [hep-ph].

\bibitem{Rizzo:1984bc} 
  T.~G.~Rizzo,
  Phys.\ Rev.\ D {\bf 32}, 43 (1985).

\bibitem{Belyaev:2012qa} 
  A.~Belyaev, N.~D.~Christensen and A.~Pukhov,
  Comput.\ Phys.\ Commun.\  {\bf 184}, 1729 (2013)
  [arXiv:1207.6082 [hep-ph]].

\bibitem{Agrawal:2014ufa} 
  P.~Agrawal, Z.~Chacko and C.~B.~Verhaaren,
  arXiv:1402.7369 [hep-ph].

\end{thebibliography}
\end{document}